\documentclass[10pt,twocolumn]{article}
\usepackage[top=1.9cm, bottom=1.9cm, left=1.8cm, right=1.8cm]{geometry}

\usepackage{amsmath,amssymb,amsfonts}
\usepackage{listings}                       %
\usepackage[noend]{algpseudocode}
\usepackage{multirow}
\usepackage{graphicx}
\usepackage{textcomp}
\usepackage{xcolor}
\usepackage{flushend}
\usepackage[lined,boxed]{algorithm2e}

\usepackage{graphicx}
\usepackage[bf]{caption}
\usepackage[square,sort,numbers]{natbib}
\SetCommentSty{mycommfont}

\usepackage{xcolor}
\usepackage{booktabs}
\usepackage{graphicx}
\usepackage{paralist}
\usepackage{subcaption}

\usepackage{times}
\renewcommand{\footnotesize}{\fontsize{9}{10}\selectfont}
\usepackage[compact]{titlesec}
\titlespacing*{\section}{0pt}{*3}{3pt}
\titlespacing{\subsection}{0pt}{*2}{2pt}
\titlespacing{\subsubsection}{0pt}{*3}{3pt}

\definecolor{linkcol}{rgb}{0,0,0.5}
\definecolor{citecol}{rgb}{0,0.5,0.3}
\definecolor{urlcol}{rgb}{0.3,0,0}

\usepackage{xspace}

\let\OLDthebibliography\thebibliography
\renewcommand\thebibliography[1]{
  \OLDthebibliography{#1}
  \setlength{\parskip}{0pt}
  \setlength{\itemsep}{0pt plus 0.3ex}
}

\usepackage[hang,flushmargin]{footmisc}

\renewcommand{\footnoterule}{%
  \kern -3pt
  \hrule width 1in
  \kern 2pt
}

\usepackage{url}
\makeatletter
\def\url@leostyle{%
  \@ifundefined{selectfont}{\def\UrlFont{}}%
  {\def\UrlFont{}}%
}
\makeatother
\usepackage[hyphenbreaks]{breakurl}

\definecolor{darkred}{RGB}{153,0,0}
\definecolor{darkblue}{RGB}{0,0,99}
\usepackage[colorlinks=true, linkcolor = darkred,   citecolor = darkred, urlcolor = darkblue]{hyperref}

\newtheorem{ldp-definition}{Definition}
\newtheorem{dp-definition}{Definition}
\newcommand{\descr}[1]{\smallskip\noindent\textbf{#1}}

\newcommand{\code}[1]{\texttt{\small{#1}}}

\title{\bf Deliberate Exposure to Opposing Views and its Association with Behavior and Rewards on Political Communities\footnote{To appear at WWW 2024, please cite accordingly. Corresponding author: alexandros.efstratiou.20@ucl.ac.uk}}

\author {
    Alexandros Efstratiou\\[1ex]
    University College London}

\date{}

\begin{document}

\maketitle

\begin{abstract}
  Engaging with diverse political views is important for reaching better collective decisions, however, users online tend to remain confined within ideologically homogeneous spaces.
  In this work, we study users who are members of these spaces but who also show a willingness to engage with diverse views, as they have the potential to introduce more informational diversity into their communities.
  Across four Reddit communities (\emph{r/Conservative, r/The\_Donald, r/ChapoTrapHouse, r/SandersForPresident}), we find that these users tend to use less hostile and more advanced and personable language, but receive fewer social rewards from their peers compared to others.
  We also find that social sanctions on the discussion community \emph{r/changemyview} are insufficient to drive them out in the short term, though they may play a role over the longer term.
\end{abstract}

\section{Introduction}\label{sec:intro}

Social media can widen democratic participation and promote information exchange~\cite{loader_networking_2011}.
However, they may also absorb users into online groups, potentially giving rise to uncivil interactions dominated by a select few users~\cite{halpern_social_2013}.
% Furthermore, they introduce other risks such as disinformation targeted towards users, which can result in societal fractures~\cite{mckay_disinformation_2021}.

Constructive or deliberative interactions between people with diverse views can lead to higher-quality information exchange, even when such interactions are competitive~\cite{semaan_designing_2015,shi_wisdom_2019,wright_democracy_2007}.
However, online interactions mostly occur with homogeneous ideas and users (echo chambers)~\cite{efstratiou_non-polar_2023,terren_echo_2021,waller_quantifying_2021,zollo_debunking_2017}, and when they do happen between users of opposing ideologies, they tend to be negative and unconstructive~\cite{bliuc_you_2020,de_francisci_morales_no_2021,marchal_be_2021}.
Moreover, political disagreements may lead users to disengage from politics~\cite{torcal_revisiting_2014} or to seek out views that reaffirm their initial beliefs~\cite{bright_how_2022,weeks_incidental_2017}.

Users who are part of such homogeneous online groups but who otherwise demonstrate a willingness to partake in heterogeneous discussions can be a promising avenue for bringing new ideas into these groups, yet this remains understudied.
Here, we set out to better understand these users and whether they tend to be penalized by their own communities, as this may limit their influence and participation.
We operationalize such users as those who post or comment in the \textit{r/changemyview} (CMV) subreddit, i.e., users who \textit{deliberately} seek out and engage with opposing views, but who are also active in political subreddits with more defined ideological alignments.
We also seek to understand whether these users are resilient to social punishments when deliberating with others.
Specifically, we pose the following research questions:

\begin{itemize}
  \item[\textbf{RQ1}] Do CMV participants receive fewer social rewards (i.e., net upvotes) in their home communities than non-participants?
  \item[\textbf{RQ2}] What are the differences in the language used between CMV participants and non-participants, if any?
  \item[\textbf{RQ3}] Can social punishments (i.e., downvotes) in discussion communities drive users out of these communities and into seclusion?
\end{itemize}

\descr{r/changemyview.}
CMV has been described as an ``anti-echo chamber''~\cite{guest_anti-echo_2018}.
Users make submissions asking other users to present arguments against an opinion they hold in the comments.
Arguments must be made genuinely, and posting users must truly be willing to change their view.
The community is heavily moderated for civility and engaging in good faith, and is in the top 150 most subscribed subreddits.\footnote{https://www.reddit.com/best/communities/1/}
We rely on this community for our methodology as it characterizes a group of users who \textit{deliberately} expose themselves to diverse views, either by inviting them or counter-arguing them, in a \textit{civil} and \textit{genuine} manner as mandated by the subreddit's rules~\cite{koshy_measuring_2023}.\footnote{https://www.reddit.com/r/changemyview/}

\descr{Methodology.}
To address \textbf{RQ1}, we obtain data from four political Reddit communities (a.k.a., \textit{subreddits}), which lie between the far left and far right of the political spectrum (\emph{r/ChapoTrapHouse; CTH, r/SandersForPresident; SFP, r/Conservative; CON, r/The\_Donald; TD}).
% This yields 48.2M comments from 799K users.
In Section~\ref{sec:dissent_analysis}, we allocate each user to one of the four subreddits based on commenting activity.
Then, from each community, we subset users who also participate in CMV, i.e., those who actively seek out opposing views.
We match them to other similar users in their community and compare the net upvotes that they receive.

For \textbf{RQ2}, we analyze linguistic differences between CMV participants and other users in Section~\ref{sec:linguistic}.
We examine their language's grade level using readability formulas, hostility as determined through Perspective API models, psychological traits using LIWC-22, and entities and topics they discuss.

For \textbf{RQ3}, we utilize a full year of data for \textit{all} users who appear in CMV (10.1M comments from 76.8K users) in Section~\ref{sec:cmv_trails}.
We obtain their user trails, looking at which subreddits they comment on and their comments' scores.
% We identify these users' ``home'' subreddits based on their activity in that year, and obtain their temporal user trails looking at \emph{where} they post each comment (CMV, home, or other), and whether this comment is upvoted, downvoted, or neutral.
Using higher-order Markov chains, we compute transition probabilities from certain communities to others, given their history's subreddits and scores.
% Furthermore, to observe longer-term patterns, we separate out users who continue engaging in CMV later into the year, then compare the proportion of downvoted comments with other users.
% Finally, we consider whether one-time commenters amass more downvotes than users who go on to post more comments in the community.
% we consider the role of downvotes on the first comment a user makes in CMV, by analyzing the proportion of downvotes among one-time commenters compared to commenters who continue to engage with the community.

\descr{Main findings.}
Overall, we find that CMV participants receive 2.53\% to 19.22\% fewer upvotes than non-participants in their home communities.
% across all four subreddits of interest.
They also differ in linguistic style, using higher-grade text, less hostile and confident but more personable and authentic language, and discussing slightly different topics. %; however, both groups of users discuss largely the same kinds of entities.

Social sanctions are not enough to drive users out of CMV in the short term, although sustained sanctions over the longer term have a less clear role.
% However, those who stay in the discussion community over the longer term attract fewer downvotes than users who eventually leave.
%, raising considerations around sustained sanctions.
% Surprisingly, one-time commenters receive \emph{fewer} downvotes than commenters who go on to post more, perhaps because the latter are motivated to make other people see their views.
% or because downvotes are a consequence of engaging with highly contentious discussions which warrant further activity.
% This pattern somewhat agrees with our findings from the short-term analysis.

Our findings have several implications.
First, though accepted by their communities, users with diversified exposure are not as popular.
Thus, harnessing their openness by making their voices more prominent within their own spaces is an open challenge.
Second, their language is more moderate, which could be linked to their lower popularity; efforts to make such language more normative over time may be fruitful.
% ; users' popularity may suffer if they are more reasonable
%, above simply attempting to immediately introduce it into political communities.
Finally, although disapproval does not drive users away in the short term, it may need to be balanced over the longer term to encourage continued engagement.

\section{Related work}\label{sec:lit_review}

In this section, we cover work on why users are socially rewarded and the role of such rewards on engagement.
Moreover, we look at the type of content that is preferred by denizens of online spaces.

\descr{Who is rewarded online?}
% Users who receive more punishments (downvotes) than rewards (upvotes) on Reddit are somewhat rare and usually exhibit troll-like features, so they are less likely to receive replies by other users~\cite{ashford_assessing_2020}.
% % Users who are consistently punished on Reddit (i.e., have net negative scores across all of their comments) are somewhat rare and their account features tend to point towards troll-like behavior, such as being more recently-created, having fewer overall comments (which may mean the accounts get banned or abandoned), and smaller gaps between each comment~\cite{ashford_assessing_2020}.
% % As a result, these users are less likely to receive replies by others~\cite{ashford_assessing_2020}.
% However, there are still differences in who receives the most social rewards even among non-troll users.
On Twitter, partisan content receives more engagement than bipartisan or neutral content~\cite{garimella_political_2018}, which suggests that more moderate voices or network ``brokers'' are penalized.
Indeed, on the far-right, pro-Trump Reddit community \textit{r/The\_Donald}, the 1000 most upvoted comments in 2017 featured substantially more extreme and hateful speech than less-upvoted comments~\cite{gaudette_upvoting_2021}.
Furthermore, interviews with moderators of the Reddit community \textit{r/AskHistorians}, which aims to provide accurate descriptions of historical events, reveal that visitors of the community tend to mostly upvote comments which seem attractive and align with their biases, while more accurate comments receive fewer upvotes~\cite{gilbert_i_2020}.
On the other end, users who eventually leave the conspiracy-minded QAnon community begin to receive lower net scores on their dissenting comments leading up to their departure~\cite{phadke_characterizing_2021}.
\citet{petruzzellis_relation_2023} report that opinion changes as a result of CMV discussions can also lead a user to leave or join other Reddit communities, suggesting that user behavior (and by extension the rewards received) may be related to the user's willingness to engage with other views.

\descr{The role of rewards on engagement.}
Users may look at their own comments' scores to gauge support from their community~\cite{jhaver_did_2019}, therefore, these scores can affect their engagement.
%  social rewards and sanctions can affect their types and degrees of engagement.
Some Reddit users express negative sentiment after being downvoted, but positive sentiment after being upvoted~\cite{davis_emotional_2021}.
Surveys reveal that comment score and user status are motivating factors behind why Reddit users may choose to participate in discussions on the platform~\cite{moore_redditors_2017}.
% Furthermore, user scores in conspiratorial and right-leaning communities on Reddit are predictive of their political preferences 4 years later~\cite{massachs_roots_2020}.
% This could be because such preferences may align with existing ideologies, or because users perceive acceptance by these communities. 

Scores may also affect engagement on social Q\&A sites, e.g., Stack Exchange.
Upvotes on answers are linked with more subsequent contributions~\cite{wei_motivating_2015}, while new users may decide whether to continue participating in such sites based on the scores that their questions receive~\cite{kang_motivational_2022}.
However, in some situations, the opposite effect holds; upvotes may reduce contributions, perhaps because users do not want to risk their good reputation, while downvotes may motivate users to improve their scores by engaging more~\cite{mustafa_what_2022}.

\descr{Current literature gaps.}
Existing work demonstrates that more neutral or disagreeable users receive fewer rewards from others~\cite{garimella_political_2018,phadke_characterizing_2021}, and how such rewards or sanctions motivate users' engagement~\cite{davis_emotional_2021,moore_redditors_2017}.
% massachs_roots_2020}.
Yet, it remains unclear whether members of communities with specific narratives but who are otherwise willing to engage with broader views through good-faith discussions are penalized by their peers.

These users are important to understand, as they may be uniquely positioned to bring more diverse ideas into their communities or normalize more open-minded language~\cite{danescu-niculescu-mizil_no_2013}.
This is especially pertinent given that people are often secluded in specific groups~\cite{zollo_debunking_2017} or turn hostile when engaging with other-minded people~\cite{de_francisci_morales_no_2021,marchal_be_2021}, thus making them apprehensive of influence from ``outsiders''~\cite{zhang_community_2017}.
At the same time, it is worth studying whether social rewards can affect engagement even in communities that are specifically designed for wide-ranging discussions (and thus run a high risk of encountering disagreement), as this may carry implications for how online deliberations are enacted.
In this paper, we aim to address both of these gaps.

\section{Dataset}\label{sec:data}

% To capture a comprehensive picture of how CMV participants are treated across different political communities, 
We obtain data from a far-left (\textit{r/ChapoTrapHouse}; CTH), a moderate-left (\textit{r/SandersForPresident}; SFP), a moderate-right (\textit{r/Conservative}; CON), and a far-right (\textit{r/The\_Donald}; TD) subreddit, using the Pushshift API~\cite{baumgartner_pushshift_2020-1}.
CTH and TD were banned in 2020 for Reddit rule violations, including promoting violence.
We collect data between July 20th, 2016, which is the creation of the youngest subreddit among the four (CTH), and December 31st, 2019.
We choose these four subreddits because preliminary analyses revealed that they are all among the top 20 political subreddits\footnote{We define a political subreddit as any subreddit in which 50\% or more of the comments are political, as determined by~\citet{rajadesingan_political_2021}.} in terms of participating users, and they all have specific accepted narratives.
TD and SFP advocate for Donald Trump and Bernie Sanders, while CON and CTH espouse conservatism and anti-capitalism, respectively.

In addition, we collect one year of data for all users who appear in CMV between January 1st and December 31st. %, 2018 (as this year was unaffected by presidential elections, campaigns, or inaugurations).
% There, users engage in good-faith discussions to change another person's view (or have their own views changed).
% , and we therefore treat this as our discussion community.
% We select 2018 as it is the only year during our sampling period where no presidential election, inauguration, or campaign was ongoing in the US, thus, we expect fewer confounding factors.
Table~\ref{tab:description} is a description of these datasets. 
%, although we perform further filtering which we explain in the relevant sections.

\begin{table}[t!]
  \centering
  \footnotesize
  \begin{tabular}{llrrr}
  \toprule
    \textbf{Dataset} & \textbf{Sub} & \textbf{\#comments} & \textbf{\#authors} & \textbf{Dates} \\
    \midrule
    \multirow{5}{*}{A} & TD & 37.7M & 545K & \multirow{5}{*}{Jul 16-Dec 19} \\
    % A & TD & 37.7M & 545K & \multirow{5}{Jun 16-Dec 19} \\
     & CTH & 7.00M & 108K & \\
     & CON & 2.17M & 140K & \\
     & SFP & 1.31M & 146K & \\
     & \textbf{total} & \textbf{48.2M} & \textbf{799K} & \\
     \midrule
    B & CMV & 10.2M & 76.9K & Jan-Dec 18 \\
    \bottomrule
  \end{tabular}
  \caption{Data description.}% Note that the \#comments reported for CMV are spread across 792 subreddits (including CMV) where users who appeared on CMV also commented in.}
  \label{tab:description}
\end{table}

\section{Penalties to CMV Participants}\label{sec:dissent_analysis}

\begin{figure}[t!]
    \centering
    \includegraphics[width=0.75\columnwidth]{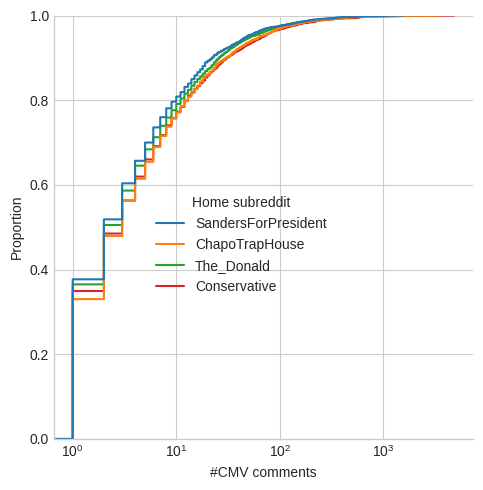}
    \caption{ECDF of number of CMV comments per participant.}
    \label{fig:cmv_ecdf}
\end{figure}

% See also Moyer et al. (page 2 section on Reddit background and terminology) for some considerations wrt using upvotes and downvotes for analyses in Reddit
In this section, we compare comment scores between users who deliberately expose themselves to opposing views (i.e., CMV participants) and others.
% Karma rewards obtained by users who are open to dialogue with other views (i.e., dissent-seeking) to those obtained by users who do not demonstrate such openness.
% Submissions must be made in good faith (i.e., users must truly be willing to change their views) and moderation enforces civil discussion.
% , and users who have their views changed as a result of a comment can reward that commenter with a ``Delta''.
% This subreddit has been referred to as an ``anti-echo chamber''~\cite{guest_anti-echo_2018}.*
We treat any user who has made at least one comment \textit{or} submission on CMV as a participant, and a non-participant otherwise.
Due to subreddit rules, users who make submissions must also comment and respond to other users under their submission.
Though it could be argued that submitters and commenters differ in their motivation for engaging, we are able to replicate our main analyses when excluding submitters.
Thus, we treat CMV participants as those with either submissions or comments in the subreddit throughout the rest of the paper.
Figure~\ref{fig:cmv_ecdf} shows the Empirical Cumulative Distribution Function (ECDF) of the number of comments made on CMV by CMV participants across the four subreddit homes.
% Although submissions and comments differ in nature, both involve good-faith discussions between different-minded people and both pertain to active engagement with opposing views.
% there are qualitative differences between users who post submissions (and invite others to change their view) and those who post comments (thus attempting to change someone else's view), we argue that all users on this community are willing to engage in good-faith discussions with people of various opinions.
%, therefore making them good candidates for the purposes of this study.
% There are potential arguments against this classification; e.g., despite its popularity, some users may simply be unaware of CMV's existence. 
%, or engage with other discussion communities.
% For this reason, we perform a validation study in Section~\ref{sec:validation}.

\subsection{User Allocation and Matching}\label{sec:matching}

Following previous work~\cite{an_political_2019,efstratiou_non-polar_2023,rajadesingan_political_2021}, we assign users to one of the four subreddits as their ``home'' if, within our data pool, they have the majority of their comments \emph{and} an overall score above 1 (Reddit's default comment score) there.
% This approach has been used in previous work~\cite{rajadesingan_political_2021,an_political_2019}.

To obtain comparable \emph{case} (i.e., CMV participant) and \emph{control} (i.e., non-participant) groups, we follow a similar matching approach to~\citet{phadke_what_2021}.
First, we subset all users per subreddit who \textit{also} appear in CMV.
This forms our four case sets.
All other home users for each subreddit are potential controls for the respective subreddit's case set.
We then match cases to potential controls on 7 features: total comments during the 1) observation and 2) pre-observation period, proportion of comments in their home subreddit during 3) observation and 4) pre-observation, total subreddits commented in during 5) observation and 6) pre-observation, and 7) date of their first comment (to the nearest day).
% We select these features as they capture the extent and type of user activity, the breadth of communities they tend to interact with, and any potential time-dependencies in such activity and interactions.
We only consider activity on political subreddits~\cite{rajadesingan_political_2021}.
% Therefore, we do not consider semantic similarity between subreddits as they all discuss the same topic.
% Moreover, we do not match users exactly on the month of their first comment as~\citet{phadke_what_2021} do.
% Instead, we take the time of posting for each user's first comment (to the nearest day) and use it as another continuous variable on which to match.
% We do this for two reasons.
% First, it can capture more similar users in terms of first comment.
% For example, a user who has posted their first comment on October 31st is temporally closer to someone who posted their first comment on November 1st than someone who posted on October 1st, although this matching would not be possible if matching exactly on contribution month.
% Secondly, it allows for a bigger pool of users to choose from for the other features, as we would need to subset our potential sample based on contribution month for ~\citet{phadke_what_2021}'s method.~\footnote{We also attempt exact month matching, however, this underperforms compared to our variable daily matching in getting more similar user sets (results in higher standardized mean differences).}

Based on these features, we conduct nearest-neighbor Mahalanobis distance matching with replacement.
%  (such that a single control user can be matched to multiple treated users).
We remove pairs which contain bots (see Appendix~\ref{app:bots}), although sensitivity analyses reveal almost identical results when including bots.
To assess matching robustness, we obtain Standardized Mean Differences (SMDs) between the case and control sets for each subreddit and across each matching feature.
SMDs below 0.20-0.25~\cite{kiciman_using_2018,phadke_what_2021,stuart_matching_2010} generally indicate good matching.
No SMD exceeds 0.15 for any feature, indicating robust matching (Table~\ref{tab:smds}).
% Overall, we retain 80.0K matched users across all subreddits (see Table~\ref{tab:smds} for the exact breakdown).

\begin{table*}[ht!]
  \centering
  \footnotesize
  \begin{tabular}{lr|rrr|rrr|rr}
  \toprule
    & & \multicolumn{6}{c}{\textbf{Standardized Mean Difference}} & & \\
    & & \multicolumn{3}{c}{\textbf{Observation}} & \multicolumn{3}{c}{\textbf{Pre-Observation}} & & \\
    \textbf{Subreddit} & \textbf{1st comment}& \textbf{\#comments} & \textbf{\%home} & \textbf{\#subs} & \textbf{\#comments} & \textbf{\%home} & \textbf{\#subs} & \textbf{\#cases} & \textbf{\#controls} \\
    \midrule
      r/The\_Donald & -0.03 & 0.05 & -0.03 & 0.11 & 0.03 & 0.02 & 0.05 & 19948 & 17640\\
      r/Conservative & -0.03 & 0.06 & 0.01 & 0.15 & 0.03 & 0.02 & 0.07 & 8854 & 7231\\
      r/SandersForPresident & -0.04 & 0.08 & 0.04 & 0.14 & 0.05 & 0.02 & 0.08 & 8026 & 6679\\
      r/ChapoTrapHouse & -0.04 & 0.05 & -0.02 & 0.11 & 0.03 & N/A & 0.07 & 6251 & 5339\\
      \bottomrule
    \end{tabular}
    \caption{SMDs for every feature across all subreddits. Notice the N/A value in home comments for the pre-observation period in r/ChapoTrapHouse because the subreddit did not exist then.}
    %  Negative SMDs indicate that values were overall lower in the cases than the control group. 
    \label{tab:smds}
\end{table*}

\subsection{Validation Study}\label{sec:validation}

Before proceeding, we validate the meaningfulness of the distinction between CMV participants and non-participants in each subreddit.
We follow a method by~\citet{garimella_political_2018}, who assess users' degree of partisanship by analyzing the political leaning of news sources that they share as labelled by~\citet{bakshy_exposure_2015}.
Since CMV participants should be less partisan (more open-minded), we expect that they will share news in a less one-sided manner.

\citet{bakshy_exposure_2015} assign model predictions between -1 (fully left) and 1 (fully right) to the top 500 domains on Facebook in the first half of 2014. 
We reinforce this with a more recent 2019 dataset~\cite{votta_favstatsallsider_2023}, which presents bias ratings for 548 sources labeled by human assessors from AllSides Media Bias.
%\footnote{https://github.com/favstats/AllSideR}
Ratings have 5 categories, from -2 (very left) to 2 (very right), with 0 indicating center.
We transform the Bakshy et al. dataset into this categorical scale by splitting the continuous scores into 6 even bins (and taking the two middle bins to be center).
% We transform the Bakshy et al. dataset into this categorical scale based on 6 bins; the two middle bins (-0.33 to 0.33) indicate center, with the remaining 4 bins in either direction being assigned their respective categorical leanings.
For domains that are in both datasets but have different scores, we keep the human-assessed AllSides score.
Excluding duplicates, we consider 795 domains.
% We find the same patterns as the ones reported below when omitting the Bakshy et al. dataset.

% We find that this dataset alone results in a notable left-leaning slant among \emph{all} groups, even for right-leaning subreddits, possibly due to its age (2014/15) or platform differences (as it presents data only for the 500 most shared websites on \emph{Facebook}).
% Thus, we also use another dataset which presents bias ratings for 548 sources, labelled by human assessors from AllSides Media Bias.\footnote{\href{https://github.com/favstats/AllSideR}{https://github.com/favstats/AllSideR}}
% Ratings have 5 categories, from -2 (left) to 2 (right), with 0 indicating center, -1 left-center, and 1 right-center.
% % and middle categories left or right slant.
% We reinforce this dataset with the data from~\citet{bakshy_exposure_2015}, which ascribes model predictions between -1 and 1 on a continuous scale.
% We transform this into a categorical scale based on 6 bins; the two middle bins (-0.33 to 0.33) indicate center, with the remaining 4 bins in either direction being assigned their respective categorical leanings.
% For domains which appear in both datasets but have different scores, we keep the score from the AllSides dataset.
% % because it is assessed using in-depth analysis by humans, as opposed to the Bakshy et al. dataset which relies on model predictions.
% We find the same patterns to the ones reported below when omitting the Bakshy et al. dataset.

We extract URLs from every comment that users post in any subreddit across the observation period, and take the mean domain score across all comments for each group per subreddit.
We also do this for fully random samples of users as an additional validation layer ($N_{random} = N_{cases}$).
% of a size equal to the number of cases we have per subreddit for comparison.
As expected, cases are less biased than controls and random users in every subreddit; that is, left-leaning cases (CTH and SFP) post less left-leaning, and right-leaning cases (CON and TD) post less right-leaning URLs than control or random users (Table~\ref{tab:media_bias}).
This also holds when measuring bias at the user, instead of the comment level.

\begin{table}[t]{}
  \centering
  \footnotesize
  \begin{tabular}{lrrrrr}
    \toprule
      \textbf{Subreddit} & $M_{case}$ & $M_{control}$ & $M_{random}$ \\
      \midrule
      TD & -0.03 & 0.32 & 0.56 \\
      CON & -0.27 & -0.01 & 0.18 \\
      SFP & -0.73 & -0.78 & -0.80 \\
      CTH & -0.79 & -0.84 & -0.86 \\
      \bottomrule
    \end{tabular}
    \caption{Mean source bias. All differences between cases and controls are significant at the 0.05 level.}
    \label{tab:media_bias}
\end{table}

Note that the vast majority of the top 12 most-visited outlets~\cite{newman_reuters_2022} receive a left-leaning score (CNN, NYT, NBC/MSNBC, WaPo, BuzzFeed, CBS, ABC, HuffPost), with only one (Fox) being right-leaning and three (NPR, USAToday, MSN) being center.
This might explain the seemingly paradoxical, near-neutral score of TD cases and CON controls, and the left-leaning score of CON cases (i.e., they likely simply share more mainstream sources than others).
Due to our user matching, controls are also more moderate than random samples.
Overall, this check validates that CMV participants are less partisan than controls or random users.

% While TD cases and CON controls seem to be nearly neutral, CON cases actually appear to share mostly left-leaning sources; this is possibly because the vast majority of the most mainstream outlets~\cite{newman_reuters_2022} receive a left-leaning score (CNN, NYT, NBC/MSNBC, WaPo, BuzzFeed, CBS, ABC, HuffPost), with only one (Fox) being right-leaning and three (NPR, USAToday, MSN) being center.
% Therefore, these dissent-seeking CON users may simply tend to share more mainstream sources than others.
% Due to our user matching, controls are also more moderate than random samples.
% Overall, this check validates our method.

\subsection{Differences in Rewards}\label{sec:reward_analysis}

Next, we compare the average comment scores (i.e., upvotes minus downvotes) between cases and controls for each subreddit.
% Having matched users across the two groups and removed bots, we then compare the overall net upvotes received per comment by users between these groups per subreddit.
For robustness, we perform both parametric (value-based) and non-parametric (rank-based) comparisons.
For non-parametric inference, we perform Mann-Whitney U tests.
In cases of violation of the equality of variances assumption (i.e., significant Levene's tests), we corroborate this with a further median test.
%, which makes no distribution assumptions.

% The median test uses chi-squared tests to check whether there are differences in the number of occurrences above and below the data's grand median across the case and control groups, and therefore makes no distribution assumptions.
We show non-parametric results in Table~\ref{tab:karma_nonparam}.
Control users have significantly higher mean ranks than case users in all subreddits meaning that they receive higher scores, except for CON, which is marginally non-significant.
% A higher mean rank indicates that members in the control group generally tend to receive higher karma scores.
% For r/The\_Donald and r/SandersForPresident, where equality of variances cannot be assumed, we corroborate these patterns with significant median tests.
However, despite statistically significant differences in 3 out of 4 subreddits, all effect sizes are very small (Cohen's \textit{d} $< 0.2$).
% , which shows the difference in the standard deviation of ranks between the two groups. 

\begin{table*}[t!]
  \centering
  \footnotesize
  \begin{tabular}{ll|rr|rr|rr|rr}
    \toprule
       & & \multicolumn{2}{c}{\textbf{Case}} & \multicolumn{2}{c}{\textbf{Control}} & \multicolumn{2}{c}{\textbf{Mann-Whitney}} & \multicolumn{2}{c}{\textbf{Median test}} \\
       \midrule
       \textbf{Subreddit} & \textbf{EoV} & \textbf{Median} & \textbf{Mean rank} & \textbf{Median} & \textbf{Mean rank} & \textbf{Cohen's \textit{d}} & \textbf{\textit{p}} & \textbf{Stat} & \textbf{\textit{p}} \\
       \midrule
         r/The\_Donald & False & 3.91 & 18375 & 4.11 & 19269 & 0.082 & $<$ 0.0001 & 41.43 & $<$ 0.0001 \\
         r/Conservative & True & 5.14 & 7979 & 5.40 & 8121 & 0.030 & 0.054 & N/A & N/A \\
         r/SandersForPresident & False & 4.00 & 7171 & 4.53 & 7571 & 0.094 & $<$ 0.0001 & 23.40 & $<$ 0.0001 \\
         r/ChapoTrapHouse & True & 9 & 5567 & 10.27 & 6064 & 0.148 & $<$ 0.0001 & N/A & N/A \\
         \bottomrule
  \end{tabular}
  \caption{Medians, mean ranks, and statistics for non-parametric tests of upvotes by group. EoV is True if the equality of variances assumption is met and False otherwise, in which case we follow the Mann-Whitney test up with a median test.}
  \label{tab:karma_nonparam}
\end{table*}

% For parametric inference, we use Welch's t-tests because equality of variances is violated in all cases.
%, making independent-samples t-tests unsuitable.
% Nonetheless, sensitivity analyses show that results are largely the same using independent-samples t-tests.
For parametric inference, 
% all groups show normality violations (significant one-sample Kolmogorov-Smirnov tests against a normal distribution), although the group sizes should be robust against these.
% % Note that normality is violated in all of our groups as determined via significant one-sample Kolmogorov-Smirnov (KS) tests against a normal distribution, however, the group sizes should be robust against violations of normality even when using parametric tests.
% Normality violations persist through log-transformations; thus, 
we perform a 1000-iteration bootstrapping t-test procedure on re-samples of our data, correcting for violations in equality of variance where necessitated by the resampling for that iteration.
We do this for both non-transformed and log-transformed versions of our data, as log-transformations bring the original power distribution closer to the t-distribution.
% , although we also conduct these tests without trimming for robustness.
% keep the non-transformed data in the parametric inference.
We show parametric results in Table~\ref{tab:karma_param}, which mostly corroborate our non-parametric tests.
Controls have significantly higher mean upvotes per comment compared to cases in all subreddits, with the exception of CON in the non-transformed analysis.
% (which was non-significant in the non-parametric tests.)
Once again, the effect sizes are small.
% These differences remain significant in sensitivity checks on the entire original samples, except CON.
% For robustness, we also conduct these tests while retaining the top 5\% values.
% The differences remain significant, with the exception of r/Conservative.

\begin{table*}[t]
  \centering
  \footnotesize
  \begin{tabular}{lrrrrr}
    \toprule
      \textbf{Sub} & $M_{case}$ (G) & $M_{ctrl} (G)$ & \%diff (G) & \textbf{\textit{t}} (log) & \textbf{\textit{d}} (log) \\
      \midrule
      TD & 6.54 (1.37) & 7.54 (1.45) & 14.20 (5.67) & **-3.06 (***-8.16) & 0.032 (0.087)\\
      CON & 9.38 (1.64) & 9.62 (1.67) & 2.53 (1.81) & -0.80 (*-1.94) & 0.013 (0.032) \\
      SFP & 8.42 (1.47) & 10.21 (1.58) & 19.22 (7.21) & ***-4.33 (***-5.78) & 0.072 (0.10) \\
      CTH & 12.25 (2.04) & 13.29 (2.19) & 8.14 (7.09) & ***-3.78 (***7.71) & 0.07 (0.149) \\
      \bottomrule
    \end{tabular}
    \caption{Overall means and parametric test statistics of upvotes by group based on 1000-iteration bootstraps. G refers to geometric means, and log refers to log-transformed analyses. *\textit{p} $\leq$ 0.05, **\textit{p} $\leq$ 0.01, ***\textit{p} $\leq$ 0.001. P-values calculated based on N iterations where $M_{case} > M_{control}$. \%diff = percentage difference. \textit{d} = Cohen's d.}
    % Equality of variances is violated in all cases, therefore Welch tests are performed. All statistics are calculated with the top 5\% of values removed. Retaining the top 5\% of values does not change our findings, except for r/Conservative which becomes non-significant. 
    \label{tab:karma_param}
\end{table*}

As an additional robustness check, we perform regression analyses with robust standard errors and Huber loss for de-weighting outliers to observe whether user type has an effect on comment scores while controlling for all matching features.
User type shows a consistently significant effect on comment scores above and beyond matching features in all four subreddits; this variable also consistently shows the strongest effect. 

Altogether, CMV participants receive fewer social rewards than non-participants.
% We firmly corroborate this for three of the subreddits, whereas the evidence is more tenuous for CON.
% Furthermore, mean and median upvotes are lower for dissent-seeking users in all subreddits.
The effect sizes are small, which suggests that cases and controls have high distribution overlap in middle values, but low overlap in extreme values at the long tails; in other words, control users possess the highest-ranked comments.
Therefore, despite the small effect sizes, controls may have a substantially higher potential for influence within their communities.
This is because extremely popular content is likely to receive even more engagement in a ``rich-get-richer'' phenomenon~\cite{hessel_cats_2017,horne_identifying_2017,stoddard_popularity_2015}, meaning that more community members will be exposed to these highly-upvoted comments.
% The effect sizes are small, but they are comparatively higher in the left-leaning than in the right-leaning subreddits.
% This is especially evident in the case of CTH.

\section{Linguistic Differences}\label{sec:linguistic}

% Given that we find consistent, albeit small, differences in the social rewards that dissent-seeking vs. non-dissent-seeking users receive on our four subreddits of interest, we 
Next, we analyze linguistic differences between CMV participants and non-participants.
% Our reasoning for this analysis is that, since we analyze activity on users' home communities, the votes they receive may be influenced by the language they use in these spaces (which may further be related to their dissent-seeking tendencies).
Specifically, we look at: 1) ease of readability, 2) hostility (toxic, profane, or insulting language), 3) psychological traits (analytical, authentic, and confident language), and 4) topics.

\subsection{Readability}

\begin{table}[t]
  \centering
  \footnotesize
  \begin{tabular}{l|rrr|rrr}
    \toprule
       & \multicolumn{3}{c}{\textbf{Case}} & \multicolumn{3}{c}{\textbf{Control}} \\
      \midrule
      \textbf{Sub} & \textbf{\#ret} & \textbf{\#total} & \textbf{\%ret} & \textbf{\#ret} & \textbf{\#total} & \textbf{\%ret} \\
        \midrule
        TD & 88.2K & 2.31M & 3.82 & 73.2K & 2.51M & 2.92 \\
        CON & 30.3K & 329K & 9.21 & 19.6K & 241K & 8.12 \\
        SFP & 18.5K & 152K & 12.14 & 9.85K & 104K & 9.48 \\
        CTH & 47.2K & 948K & 4.98 & 40.0K & 919K & 4.34 \\
        \midrule
        \textbf{ovrl} & \textbf{184K} & \textbf{3.74M} & \textbf{4.91} & \textbf{143K} & \textbf{3.77M} & \textbf{3.79} \\
        \bottomrule
    \end{tabular}
    \caption{Absolute (\#) and relative (\%) numbers of retained (ret) texts ($\geq$ 100 words) for readability analysis.}
    % \#ret is the absolute number of texts with > 100 words, and \#total is the total number of texts for that group. \%ret is the relative percentage of retained texts.
    \label{tab:long_texts}
\end{table}

\begin{table*}[t]
  \centering
  \footnotesize
  \begin{tabular}{ll|rr|rr|rr|rr}
    \toprule
       & & \multicolumn{2}{c}{\textbf{Case}} & \multicolumn{2}{c}{\textbf{Control}} & \multicolumn{2}{c}{\textbf{Mann-Whitney}} & \multicolumn{2}{c}{\textbf{Median test}} \\
       \midrule
       \textbf{Subreddit} & \textbf{EoV} & \textbf{Median} & \textbf{Mean rank} & \textbf{Median} & \textbf{Mean rank} & \textbf{Cohen's \textit{d}} & \textbf{\textit{p}} & \textbf{Stat} & \textbf{\textit{p}} \\
       \midrule
         r/The\_Donald & True & 9 & 80966 & 9 & 80412 & 0.012 & 0.016 & N/A & N/A \\
         r/Conservative & False & 9 & 25405 & 9 & 24211 & 0.081 & $<$ 0.0001 & 59.31 & $<$ 0.0001 \\
         r/SandersForPresident & False & 9 & 14441 & 9 & 13629 & 0.095 & $<$ 0.0001 & 39.17 & $<$ 0.0001 \\
         r/ChapoTrapHouse & False & 11 & 44276 & 10 & 42682 & 0.063 & $<$ 0.0001 & 57.33 & $<$ 0.0001 \\
         \bottomrule
  \end{tabular}
  \caption{Medians, mean ranks, and statistics for non-parametric tests of readability by group. EoV is True if the equality of variances assumption is met and False otherwise, in which case we follow the Mann-Whitney test up with a median test.}
  \label{tab:readability}
\end{table*}

% Readability refers to how easy language in a text is to understand.
Readability tests return a US school grade level (e.g., 4, 5), denoting that the text would be easily understood by an average 4th, 5th, etc. grader.
Most require at least 100 words per text, thus, we filter out comments with fewer words than this; we show how many are retained in Table~\ref{tab:long_texts}.
% We filter out texts containing fewer than 100 words for this analysis due to the readability tests' reliance on number of characters, syllables, or words, which necessitates at least 100 words per text.
% Thus, this analysis only concerns a small subset of comments.
% The non-extreme subreddits (CON and SFP) have higher relative percentages (but lower absolute numbers) of long texts.
%, but lower absolute values since these subreddits have fewer total comments.
In all subreddits, case users post more long texts than control users, both in absolute and relative terms. 
% Overall, we only analyze a small subset of comments for readability (327K out of 7.51M).

Using Python's \code{textstat} library, we obtain a readability grade level for each comment using 8 different formulas (see Appendix~\ref{app:formulas}).
We then take the modal grade level across all formulas for each comment.
%, reflecting the most agreed-upon readability level.
As this data is ordinal, we use Mann-Whitney U tests to assess differences between the ranks of cases and controls, with further median tests where equality of variances is violated.
% Once again, we corroborate these using median tests in cases where equality of variances is violated.
% As this data is ordinal, we do not perform any parametric tests.

The mean rank of cases is significantly higher than the mean rank of controls for all four subreddits, suggesting that comments made by CMV participants are overall more difficult to read in terms of grade level (Table~\ref{tab:readability}).
Again, all effect sizes are fairly small, but they are comparatively larger for the non-extreme subreddits (CON and SFP).
% We also observe similar patterns as the upvotes analysis with respect to the effect sizes being fairly small.
% CTH users use more advanced language (median grades of 11 and 10 for case and control groups, respectively) compared to other subreddits (grade 9 for all others).
% In any case, dissent-seeking users tend to produce higher-difficulty text in terms of readability than non-dissent-seeking ones across all 4 subreddits, although this difference is very small.

\subsection{Hostility Attributes}

Our next analysis focuses on whether non-participants use more hostile language than CMV participants.
Specifically, we examine whether there are significant differences in the proportion of toxic, insulting, or profane comments between case and control groups for each subreddit.
To that end, we use three models from Google Jigsaw's Perspective API~\cite{jigsaw_about_2023}:
\begin{enumerate}
  \item {\em Severe Toxicity}, defined as ``a very hateful, aggressive, disrespectful comment or otherwise very likely to make a user leave a discussion or withhold their perspective.''
  \item {\em Insult}, defined as an ``insulting, inflammatory, or negative comment towards a person or a group of people.''
  \item {\em Profanity}, defined as ``swear words, curse words, or other obscene or profane language.''
\end{enumerate}

Although Perspective is sensitive to adversarial text~\cite{hosseini_deceiving_2017}, it outperforms alternative models~\cite{zannettou_measuring_2020} and has been found to be suitable for Reddit content~\cite{chong_understanding_2022,rajadesingan_quick_2020,xia_exploring_2020}.
The models return values ranging from 0 to 1.
% which reflects the percentage of people that would find that the comment has the respective attribute.
% , with 1 indicating absolute certainty that the given attribute is demonstrated in the comment, 0 indicating absolute certainty that the attribute is absent from the comment, and 0.5 indicating complete uncertainty.
We classify a text as having the respective attribute if it has a score of $\geq 0.8$ for all models, which is adequately high to avoid false positives~\cite{hoseini_globalization_2023,kumar_designing_2021}.
% Furthermore, we use the Severe Toxicity instead of Perspective's regular Toxicity model, as the former is more robust against positive uses of curse words; we capture curse words of any kind using the Profanity model.

With this classification, we obtain one contingency table per attribute for each subreddit (12 tables in total), on each of which we perform chi-square tests (Table~\ref{tab:attributes}).
% We then perform a chi-square test on each table and show the results in Table~\ref{tab:attributes}.
% , which compares observed frequencies against {\em expected} frequencies estimated based on the overall ratio of attributed vs. non-attributed comments.
% Therefore, these tests allow us to infer whether case or control groups have a significantly higher (or lower) number of attribute-laden comments than expected.
We apply Bonferroni corrections as we make 3 comparisons with each population; thus, we interpret significance at \textit{p} = 0.017.

\begin{table}[t]
  \centering
  \footnotesize
  \begin{tabular}{clrr}
  \toprule
    \textbf{Sub} & \textbf{Attr} & $\chi^2$ & $\uparrow$\textbf{group} \\
      \midrule
      \multirow{3}{*}{CTH} & Insult & ***22.39 & Control \\
       & Profanity & **14.64 & Case \\
       & Toxicity & 1.43 & N/A \\
      \midrule
      \multirow{3}{*}{CON} & Insult & ***1558.71 & Control \\
       & Profanity & ***685.90 & Control \\
       & Toxicity & ***153.55 & Control \\
      \midrule
      \multirow{3}{*}{SFP} & Insult & 0.06 & N/A \\
        & Profanity & **9.39 & Control \\
        & Toxicity & ***24.04 & Control \\
      \midrule
      \multirow{3}{*}{TD} & Insult & ***490.49 & Control \\
        & Profanity & ***684.63 & Control \\
        & Toxicity & ***725.52 & Control \\
    \bottomrule
  \end{tabular}
  \caption{Chi-square results for Perspective attributes per subreddit. ***\textit{p} $<$ 0.0001, **\textit{p} $<$ 0.017, *\textit{p} $<$ 0.05. $\uparrow$ indicates which group has a higher observed frequency of the attribute compared to the expected frequency. N/A = non-significant.}
  % The two-star p-value indicates the cutoff of Bonferroni correction, where we adjust $\alpha$ by dividing the 0.05 cutoff by 3 (the number of comparisons we make per subreddit). Cases which are not significant even at the 0.05 level are marked as N/A.
  % as no group is significantly more frequent than the other.}
  \label{tab:attributes}
\end{table}

In 9 out of 12 comparisons, we detect significantly more hostile language in controls compared to cases.
% in the control user texts than in the case texts.
The opposite holds for profanity in CTH; CMV participants in this subreddit use more swear words.
% The only case where we observe more hostile language from the case group is with regards to profanity in CTH, which suggests that 
CON's and TD's controls are more hostile than cases in all attributes and also show the biggest differences overall.
Every attribute is more frequent among control users in 3 out of 4 subreddits.

An exception is that CTH cases use more profanity than controls.
CTH is the most profane subreddit out of the four, with 14.7\% of all comments across both groups of users being flagged for profanity.
Based on a manual inspection, this profanity is often used in a non-malicious manner.
Therefore, this finding may be due to swear words being a fairly normative method of communication within this particular subreddit.

% So far, we observe that dissent-seeking users use a) more advanced language readability-wise, and b) less hostile language.
% Overall, it can be inferred that dissent-seeking users use more sophisticated language than non-dissent-seeking users, although we once again stress the limited size of these effects.
% In the next subsections, we attempt to understand the actual text that appears in users' comments in more depth, by analyzing the entities and topics that they discuss.

\subsection{Psychological Traits in Language}

Next, we examine psychological dimensions in users' language using three of LIWC-22's~\cite{boyd_development_2022} summary metrics (Analytic, Authentic, Clout); LIWC has been used with Reddit data in prior work~\cite{an_political_2019,horne_identifying_2017,mensah_characterizing_2020}.
Analytic reflects formal and logical language; Authentic reflects the degree to which the user avoids adjusting language to fit their social environment; and Clout reflects the confidence and social status expressed in the user's writing~\cite{pennebaker_development_2015}.
As LIWC relies on a dictionary approach, we filter out comments with fewer than 10 words in these analyses.
For robustness, we perform both parametric and non-parametric comparisons of text scores between cases and controls for all three traits, which we show in Table~\ref{tab:liwc}.

\begin{table*}
    \footnotesize
    \centering
    \begin{tabular}{lrrlrrrrrr}
        \toprule
        \textbf{sub} & \textbf{\#cases} & \textbf{\#controls} & \textbf{trait} & $M_{case}$ & $M_{control}$ & $Mdn_{case}$ & $Mdn_{control}$ & \textbf{\textit{t (d)}} & $\Delta M_{rank}$ \textbf{\textit{(d)}} \\
        \midrule
        \multirow{3}{*}{TD} & \multirow{3}{*}{1.45M} & \multirow{3}{*}{1.51M} & Analytic & 42.65 & 43.97 & 38.91 & 39.7 & ***-34.21 (0.04) & ***-32.9K (0.038) \\
         & & & Authentic & 38.43 & 37.85 & 26.78 & 24.32 & ***13.87 (0.016) & ***18.7K (0.022) \\
         & & & Clout & 47.87 & 48.96 & 40.06 & 40.06 & ***-25.94 (0.03) & ***-25.9K (0.03) \\
        \midrule
        \multirow{3}{*}{CTH} & \multirow{3}{*}{609K} & \multirow{3}{*}{578K} & Analytic & 43.97 & 45.39 & 39.7 & 42.89 & ***-23.21 (0.043) & ***-14.4K (0.042) \\
         & & & Authentic & 39.9 & 39.74 & 30.98 & 30.98 & **2.43 (0.004) & **2.11K (0.006) \\
         & & & Clout & 43.93 & 44.18 & 40.06 & 40.06 & **-3.72 (0.007) & ***-2.47K (0.007) \\
        \midrule
        \multirow{3}{*}{CON} & \multirow{3}{*}{251K} & \multirow{3}{*}{185K} & Analytic & 42.3 & 43.19 & 38.85 & 39.7 & ***-9.1 (0.028) & ***-3.5K (0.028) \\
         & & & Authentic & 39.65 & 38.46 & 30.98 & 28.56 & ***11.13 (0.034) & ***4.84K (0.038) \\
         & & & Clout & 45.33 & 47.05 & 40.06 & 40.06 & ***-16.18 (0.05) & ***-6.17K (0.048) \\
        \midrule
        \multirow{3}{*}{SFP} & \multirow{3}{*}{123K} & \multirow{3}{*}{79.6K} & Analytic & 41.57 & 43.34 & 37.99 & 39.7 & ***-12.42 (0.056) & ***-3.23K (0.054) \\
         & & & Authentic & 38.41 & 38.85 & 29.13 & 28.56 & 0.68 (0.003) & **700 (0.012) \\
         & & & Clout & 45.52 & 46.39 & 40.06 & 40.06 & ***-5.58 (0.025) & ***-1.38K (0.023) \\ 
        \bottomrule
    \end{tabular}
    \caption{Descriptive and inferential statistics for LIWC analyses. Where equality of variances is violated, \textit{t} is obtained using a Welch test. \#cases and \#controls refer to the number of comments retained in analyses per group. $\Delta M_{rank}$ = difference in mean ranks used in non-parametric Mann-Whitney U tests. ***\textit{p} $<$ 0.0001, **\textit{p} $<$ 0.017, *\textit{p} $<$ 0.05.}
    \label{tab:liwc}
\end{table*}

Once again, we find consistent patterns across all four subreddits.
As expected, control users are higher in Clout (all \textit{p} $<$ 0.001), which indicates that they express more confidence and social status in their comments.
However, contrary to our expectations, control users are also higher in their use of Analytic language (all \textit{p} $<$ 0.001).
At the same time, case users use more Authentic language (all significant at the Bonferroni-corrected \textit{p} $<$ 0.017 cutoff, with the exception of SFP's parametric corroboration).

Taken together, the LIWC analysis reveals that CMV participants may not adjust their language to suit their social environment as much as non-participants do; this could partly explain the comparably fewer social rewards they garner, as self-monitoring in this way is associated with better impression management and likeability~\cite{turnley_achieving_2001}.
At the same time, non-participants write with more confidence, which can also lead to more positive evaluations by others~\cite{markowitz_self-presentation_2023}.
Though we expected CMV participants to also use more analytical language, which demonstrates more logical thinking patterns, due to their engagement in formal argumentation, we observe the opposite effect.
Based on a manual inspection of the data, this may be because texts higher in this trait tend to also exhibit more confidence, something that case users are generally lower on; for context, we provide representative texts per LIWC category and subreddit in Appendix~\ref{app:liwc}.
Thus, CMV participants may use more personable language compared to the more formal language of non-participants, perhaps because formal language is more normative within the political spaces we study.

% In order to get more context around these discussions, our next analysis focuses on the topics mentioned.

\subsection{Topic Extraction}

Next, we compare the topics that cases and controls discuss in each subreddit using Latent Dirichlet Allocation (LDA)~\cite{blei_latent_2003}.
We remove URLs and stopwords, lemmatize the text, extract bigrams, and apply Term Frequency-Inverse Document Frequency weights.
% perform text pre-processing (URL removal, stopword removal, lemmatization, tokenization), as well as bigram extraction.
Then, we iterate the number of topics hyperparameter from 5 to 15 and extract the number which produces the highest coherence score for each group (ranging from 7 to 15 topics).
% We extract the number of topics which produces the highest coherence for each group, therefore, we obtain different numbers of topics for different groups (between 8 and 15).
% Obtained coherence scores range from 0.42 (CTH controls) to 0.51 (TD cases); the lowest coherence score we obtained during testing was 0.34 (CTH controls with 6 topics).
Due to space constraints, we only show our \textit{interpretations} for the top 10 topics extracted per group in Table~\ref{tab:topics}.
We present the full topics (and constituent words) in a Google Sheet.\footnote{https://docs.google.com/spreadsheets/d/e/2PACX-1vRvanN-nTJ-DesPWxLmk56OjAOoQFNUPTKccFx35dobIGaEVjzstUzUt\\9ae2XaNEA/pubhtml}

The topic analysis affords a more grounded understanding of users' discussions.
In CTH, cases do not stray too far from political subjects.
Contrasting, controls additionally discuss art forms like movies and podcasts,
% (including other controversial left-associated podcasts)
while also hinting towards emotional states like disdain towards critics.
SFP users are more expressly political, although control users also veer into more abstract concepts such as voicing doubts and corruption concerns mostly about the right.
In TD, topics often concern ``others''; however, ``others'' mostly mean political opponents with case users (e.g., topic 5 which refers to Hilary Clinton), while they refer to other religions and countries with control users (e.g., topics 5 and 8).
With CON, we see that case users pick up on more narrow conservative talking points (abortion, guns, justice system), whereas controls adopt a more general view (e.g., right vs. wrong in topic 3 and conservative values in topic 7) with the exception of topic 6 on immigration specifically.
TD and CTH topics feature substantially more profanity than the more moderate subreddits.
Overall, differences between case and control users range from an explicit topical focus on politics to the topics' level of abstraction.
In Appendix~\ref{app:ner}, we also analyze the named entities discussed by cases and controls.
However, we do not find substantial differences in the entities or types of entities mentioned.

\begin{table*}[t]
    \centering
    \footnotesize
    \begin{tabular}{p{0.5cm}lllllllllll}
        \toprule
            \textbf{sub} & \textbf{group} & \textbf{topic 0} & \textbf{topic 1} & \textbf{topic 2} & \textbf{topic 3} & \textbf{topic 4} & \textbf{topic 5} & \textbf{topic 6} & \textbf{topic 7} & \textbf{topic 8} & \textbf{topic 9} \\
            \midrule
            \multirow{2}{*}{\textbf{CTH}} & \textbf{case} & \textit{und} & \textit{und} & voting & far left & upset & liberals & politics & israel & society & zizek \\
             & \textbf{control} & \textit{und} & \textit{und} & economy & obama & \textit{und} & disdain & america & podcasts & movies & mockery \\
            \midrule
            \multirow{2}{*}{\textbf{SFP}} & \textbf{case} & sanders & posting & medcare & election & ideology & right & russia & -- & -- & -- \\
             & \textbf{control} & sanders & primary & election & medcare & socialism & voting & companies & right & doubt & corruption \\
            \midrule
            \multirow{2}{*}{\textbf{TD}} & \textbf{case} & \textit{und} & posting & race & america & voting & emails & corruption & govt & trump & election \\
             & \textbf{control} & \textit{und} & \textit{und} & memes & election & mockery & islam & trumpism & opposition & non-us & mobilize \\
            \midrule
            \multirow{2}{*}{\textbf{CON}} & \textbf{case} & \textit{und} & govt & abortion & voting & politics & internet & reddit & guns & justice & media \\
             & \textbf{control} & voting & govt & taxes & morality & satire & media & us border & con values & -- & -- \\
            \bottomrule
    \end{tabular}
    \caption{Topic interpretations by group. \textit{Und} stands for undefined; most of these topics reflect colloquial exchanges, i.e., vague words (e.g., thanks, please, good) and/or profanity. ``Politics'' = references to \textit{both} the left and right. ``Posting'' = online activity. ``Right'' refers to the \textit{political} right.}
    \label{tab:topics}
\end{table*}

\subsection{Inferences from Linguistic Analyses}

In this section, we demonstrate that CMV participants use more moderate language compared to their peers.
Based on readability analyses, they use more advanced language, while also being less hostile in their communications as examined through the Perspective API.
LIWC analyses reveal that they are friendlier and more authentic, but less formal and confident.
The topics they discuss are seemingly less abstract and more explicitly political.
While these language aspects may be influenced from one another (e.g., the topics that CMV participants choose to engage in may necessitate more moderate language), we nonetheless provide an in-depth linguistic profile of these users, which could be a potential factor behind the fewer social rewards they accumulate.

% Overall, these analyses show that CMV participants use slightly more advanced and less hostile language, while also being friendlier and more authentic; non-participants are more formal and confident.
% While differences in entities discussed are miniscule, CMV participants' topics seem to be less abstract and more explicitly political.
% taken together demonstrate that dissent-seeking users tend to be more ``sophisticated'', although these differences are small and subtle.
% They write with higher-grade readability levels, use less hostile language, discuss \emph{slightly} more formal types of entities, and discuss more serious topics.

% However, one potentially puzzling finding is with respect to CTH; dissent-seeking and non-dissent seeking users in this subreddit show more linguistic homogeneity than the other subreddits.
% For example, case users are more profane than control users (although less insulting) with no differences in their toxicity levels and are much closer in terms of the seriousness of topics they discuss, with control users even discussing arguably more serious topics.
% Simultaneously, this subreddit also shows the largest effect size in the difference in net upvotes received by case and control users.
% This suggests that other factors or linguistic features which we have not explored may be important drivers of such differences in scores.
% Nonetheless, our linguistic analyses confirm that case users use more sophisticated language than control users.

\section{Persistence Against Downvotes}\label{sec:cmv_trails}

Next, we address whether downvotes can drive users out of discussion communities to seek approval elsewhere, e.g., in their home communities.
We obtain all comments between January 1st and December 31st, 2018, for any user who posted at least one comment in CMV during this period (excluding bots).  % utilize a different dataset using {\em any} authors who posted at least one comment in the r/changemyview community at any time between January 1st and December 31st, 2018.
This yields 10.1M comments made by 76.8K authors across 792 subreddits (including CMV).
% We select 2018 because no election, inauguration, or presidential campaign was going on in the US during that year, thus, we expect fewer external confounds.
% it was the year in which we anticipated the fewest shocks from external events; in any other year for which we had data there was either an election, inauguration, or presidential campaign going on in the US (which the majority of political subreddits concern).
%, thus this year may capture more general patterns.

% We formulate our question as a probability problem: do downvotes in a given community increase the probability that a user will leave that community/seek out a different community?
% ; that is, does the probability of leaving a discussion community/re-entering a home community increase when the user is downvoted in the discussion community?
% To that end, we build temporally sorted user trails for each of the authors in the dataset.
% , where their comments are sorted in temporal order (from earliest to latest).
We build trails for every user and split these into separate sessions when over 8 hours elapse between two successive comments.
% the time gap between one comment and the next is over 8 hours long.
%, as we assume that the subreddit of each new session may be unaffected by the previous comment's score.
For robustness, we also split at 4, 12, and $\infty$ hours (i.e., no splits). 
We note each comment's subreddit (CMV, home, or other) and score (downvoted, upvoted, or neutral).
``Home'' here means the top 3 subreddits where the user has most of their comments, as we find that \textasciitilde99\% of users have at least 50\% of their comments in $\leq$ 3 subreddits. 
% (i.e., a score of 1, which is the default score for newly posted comments).

This creates 9 possible subreddit-vote combinations, which we treat as {\em states} in a higher-order Markov chain. 
% To determine the optimal order for the Markov chain,
Between 1 - 8, we find the lowest (optimal) Bayesian Information Criterion (BIC) at the 4th order.
% We compute the Bayesian Information Criterion (BIC) from 1st to 8th order, and find an optimal (lowest) BIC at the 4th order.
% (with the 8-hour-interrupted trails).
However, we also conduct analyses at the 3rd order as a sensitivity check.
% because several possible histories never or rarely occur
% (this is also the suggested optimal order when using uninterrupted trails).
% or occur only once giving rise to extreme probabilities (0 or 1), 
We only retain user trails of size $\geq N+1$, where \emph{N = order}.

\begin{figure}[t!]
    \centering
    \includegraphics[width=0.80\columnwidth]{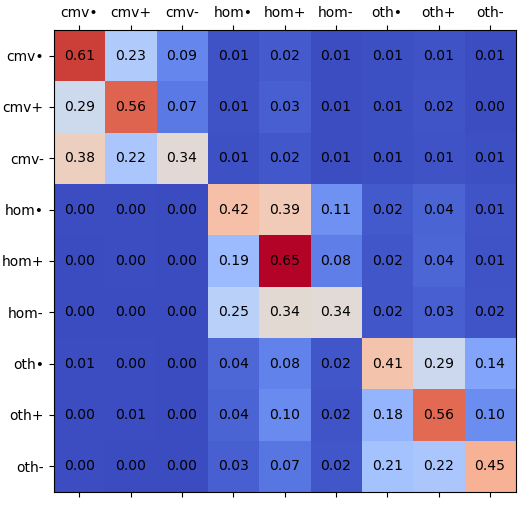}
    \caption{First-order transition matrix with 8-hour interruptions. +, -, and $\bullet$ show upvoted, downvoted, and neutral comments, respectively.}
    \label{fig:firstord}
\end{figure}

% The transition matrices reveal that, 
Regardless of score, users are always most likely to stay within their current community.
% From the transition matrices, the best predictor of the next community on which the user posts seems to be the community of the historical states; comment scores seem to play a much smaller role, if any.
For example, in the 3rd-order chain, the 27 histories with the highest probability of resulting in any given community are the 27 possible combinations of comment votes {\em in that community}, which yield recurrence probabilities between 91.1\% and 97.8\%.
% of staying in the same community.
In fact, 3 consecutive downvotes result in 0.4\%, 0.5\%, and 2.6\% \textit{higher} probabilities of staying in CMV, home, and other, respectively, than 3 consecutive upvotes.
% In fact, 3 consecutive downvotes result in slightly \emph{higher} probabilities of staying than 3 consecutive upvotes (differences of 0.4\%, 0.5\%, and 2.6\% in CMV, home, and other, respectively).
Figure~\ref{fig:firstord}, which is the first-order transition matrix, demonstrates this tendency.

\begin{figure*}[t!]
    \centering
    \includegraphics[width=0.99\textwidth]{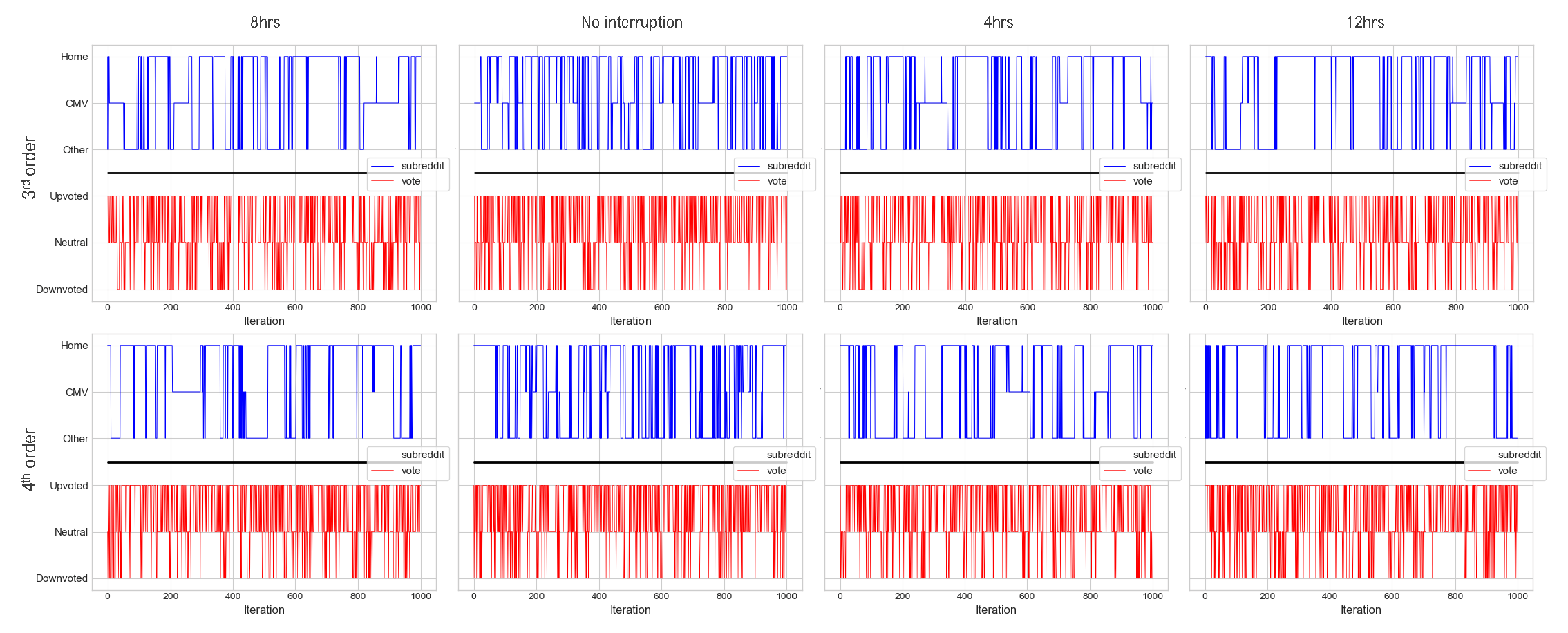}
    \caption{1000-comment Markov chain simulations with all configurations. Community fluctuations are higher with no interruption as each session's starting community is picked randomly. Unsurprisingly, users spend most of their time at home.}
    \label{fig:markov}
\end{figure*}

This pattern is also reflected in Figure~\ref{fig:markov}, where we simulate the average user's trail over 1000 comments starting with a random history and plot the ``decomposed'' resulting states.
% , separating the community and votes of the comment.
While votes highly fluctuate, there are no commensurate (lagged) fluctuations for communities.
The long flat periods indicate that commenters tend to stay in the same community, regardless of votes.
% Unsurprisingly, the simulated user spends most of their time in their home communities.
% As expected, there are more community fluctuations when no session interruption is applied, since each new session's starting community may be picked randomly.
% the starting community with each session may be picked somewhat randomly, but is treated as a transition in this case.

Overall, this analysis suggests that disagreement or social punishments are not enough to drive users away from conversations.
Instead, users mostly ``stand their ground.''
This seems to apply even when a user's very first comment in the community is downvoted (see Appendix~\ref{app:early}).
However, downvotes may still play a partial role over the longer term (see Appendix~\ref{app:longterm}).
% Although these patterns could be driven in part by conversations getting out of hand and turning increasingly hostile which may drive more engagement, the concurrent fluctuation in comment scores diminishes this explanation.
%  (i.e., the fact that they are not consistently positive or negative)
% Comment scores may not have enough time to be finalized or to substantially affect the user before the next comment, therefore, this analysis cannot rule out that consistent downvoting can ultimately drive a user out of these communities.

\section{Discussion and conclusion}

In this work, we provide a deeper understanding of users who expose themselves to diverse views.
These users are a potential avenue for introducing new ideas into communities with established narratives, therefore, we argue that they are important to study.
Specifically, we examine their treatment (sanctions and rewards) by their own communities, how this may be related to the language they use, as well as whether sanctions and rewards play a role in their engagement with discussion communities themselves.

\subsection{Implications}

Here, we reiterate our findings and explain their implications.

\descr{Rewarding bias.}
To answer \textbf{RQ1}, CMV participants receive fewer social rewards than non-participants, which contextualizes previous findings around users' preference for extremity~\cite{gaudette_upvoting_2021} and partisanship~\cite{garimella_political_2018}.
% This further contextualizes previous findings that the most upvoted comments in certain communities are also the most extreme~\cite{gaudette_upvoting_2021}, or that more partisan users receive more attention from their peers than neutral ones~\cite{garimella_political_2018}.
This suggests that communities prefer users who fully comply with established narratives, as their comments may more adequately satisfy the wider communities' biases~\cite{gilbert_i_2020}.

It is important to reiterate that CMV participants are not \emph{punished} by their communities, but rather, simply receive fewer rewards.
% Furthermore, this difference is fairly small.
% This means that they may have some influence in their communities, but perhaps not to the extent to which more secluded users can set the norms there.
However, Reddit post popularity operates on ``rich-get-richer'' mechanisms~\cite{hessel_cats_2017,horne_identifying_2017,stoddard_popularity_2015}, where upvotes result in more exposure, more upvotes, and so on. 
This could mean that CMV participants are less able to influence their communities' norms compared to their peers.
Thus, a potential problem that de-polarization scholars can examine is not the ostracization of these users per se, but rather, finding ways of making their (already accepted) voices more influential within their communities.

\descr{Costs of being moderate.}
With regards to \textbf{RQ2}, CMV participants' language is more personable and advanced and less hostile and confident, with less abstract topical foci.
% while the entities they discuss are more formal and their topics more serious.
Given that extremity is rewarded in some communities~\cite{gaudette_upvoting_2021}, this might mean that more moderate and friendlier language puts users at a disadvantage in receiving social rewards.
Thus, a potential risk is that users may be motivated to be more extreme and appear more confident in order to receive more approval from their peers, which can harm the quality of discussions taking place within the community.
% , which would agree with findings around rewarding extremity~\cite{gaudette_upvoting_2021}.

%Nonetheless, there are possibly other content or user factors which further explain this score difference; the CTH subreddit shows the largest score differences between CMV participants and other users, but smaller differences with respect to the hostility and the seriousness of topics that these two groups discuss.
Overall, this presents a challenge in that it may be moderation itself that attracts fewer social rewards, but once again, it is not \emph{penalized} per se.
In light of this, it may be more fruitful to examine pathways of making such language normative over the longer term, rather than immediately attempting to introduce it within communities.

\descr{Social approval in discussion communities.}
For \textbf{RQ3}, users mostly tend to stay in their current communities in the short term, regardless of their comments' scores.
% continue to comment on the communities they are already on in the short term,.
% we do not find any evidence that downvotes can drive users out of discussion communities in the short term; rather, .
However, it is unclear whether these scores have an effect over the longer term.
%, both in absolute terms and compared to the scores received at their homes.
% Optimistically, first-time commenters who receive downvotes are not thwarted and continue to engage with the discussion community.

% Taken together, these findings are somewhat optimistic; particularly for the shorter term, 
These findings show that users are keen to ``stand their ground,'' which is generally optimistic; CMV participants are, at least in the short term, resilient to disapproval of their views, which is an important aspect of online deliberation.
Discussions between such users are unlikely to be cut short simply due to perceived disapproval, which may allow the conversing parties to adopt new views.
This is antithetical to findings from prior work~\cite{bright_how_2022,kang_motivational_2022}, although this may be expected as we are specifically focusing on a community where disagreement is welcomed.
% On the other hand, continued engagement after downvotes could simply mean that some users engage in very heated discussions which prompt further engagement, although our Markov modeling does not reveal any substantial periods of sustained downvoting, which diminishes this explanation.

% In the long term, although users who eventually leave receive less approval from the wider community, there are possibly other, more important determinants of this departure that remain to be unpacked. 

\subsection{Limitations and Future Work}

Our work is not free from some limitations which we discuss here, along with how these could be rectified in future work.

\descr{Causality and confounders.}
As we observe naturally-occurring phenomena, we cannot ascribe a direction to these effects, e.g., whether it is users' exposure to opposing views that cause their comment scores/sophisticated language or vice-versa, or if it is a third factor driving these patterns.
% Moreover, it is worth researching whether CMV participants use more sophisticated language as a result of their exposure to diverse opinions, or if it is this sophistication which leads them to pursue more diversity of opinion.
Moreover, inaccessible deleted content in the data could warp our findings.
Strictly controlled experiments, established precedence of events, or user-reported data alongside digital trace data may be required in future work.
% To establish this, more strictly controlled experimental approaches or at least establishing precedence of events may be required in future work.

\descr{User open-mindedness.}
Though CMV participants are likely to be open-minded, this does not \textit{necessarily} mean that non-participants are the opposite.
% Therefore, some users whom we treated as non-participants may have otherwise been more open to new ideas than other users in the control group.
% Though we chose communities with fairly strict narratives to minimize the risk of this,
Future work could employ other methods of classification; for example, overall bias in the news links that users provide~\cite{garimella_political_2018} (keeping in mind that some links may be shared disingenuously), or detecting and studying more discussion subreddits.

\descr{What kind of engagement?}
For our user trail analyses, we were mostly concerned with \emph{whether} the users continued to engage in CMV, but not the \emph{quality} of such discussions.
Future work could explore this further, by examining whether users may become more subjective, hostile, or negative following a downvote even within these otherwise open-minded communities. 

% \descr{Potential confounders.}
% Several factors may affect comment scores.
% For example, individuals with lower self-esteem may be motivated to specifically seek out social rewards~\cite{scissors_whats_2016}.
% Moreover, dissent-seeking users may simply elect to engage with threads that do not attract as much engagement, thus limiting other users' exposure to them.
% Though we did not consider them here, future work could employ both digital trace and user-reported data to examine the potential role of such factors. 

\section{Ethical Statement}
This work received ethical approval from UCL's Research Ethics Committee (Project ID: 19379/001).
We only use data that is in the public domain, and do not attempt to de-anonymize or track users.
% The differences in named entities between case and control groups are minimal.
% All cells show one, two, or zero differences in the top 10 named entities between cases and controls for the 10 entity types we show (with the exception of works of art in SFP, where there are four differences.)
% The two extreme subreddits (CTH and TD) show somewhat more case-control homogeneity than the non-extreme ones; TD is the most homogeneous (5 total differences), while SFP is the least homogeneous (16 total differences).

\section*{Acknowledgments}
The author would like to thank Emiliano De Cristofaro and Christos Perivolaropoulos for their feedback on earlier drafts, and the anonymous reviewers for their helpful comments which substantially improved the paper.
This work was partially funded by the UK EPSRC grant EP/S022503/1, which supports the UCL Centre for Doctoral Training in Cybersecurity.
Any opinions, findings, conclusions, or recommendations expressed in this work are those of the author and do not necessarily reflect the views of the funder.

\small
\bibliographystyle{abbrvnat}
\bibliography{references}

\appendix

\section{Bot Removal}\label{app:bots}

\begin{figure}[t!]
    \centering
    \includegraphics[width=0.80\columnwidth]{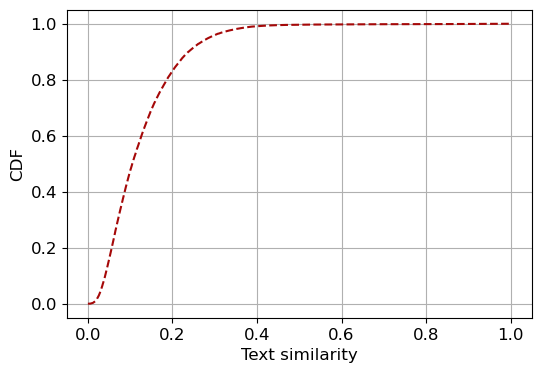}
    \caption{CDF of average comment similarity across users.}
    \label{fig:cdf}
\end{figure}

To detect bots, we obtain all of the comments posted by each of the users in our matched pairs with at least 10 comments during the observation period.
We then compute the pairwise cosine similarity between the bag-of-word vectors of each comment for each user, such that every user has their own similarity matrix.
Other than tokenization, we do \textit{not} pre-process the text to retain the exact tokens that bots repeat in their comments. 
% We do \emph{not} perform any text preprocessing (e.g., stopword or punctuation removal, lemmatization, etc.) whatsoever, other than tokenization, because most bots only change a few variable words between comments while keeping everything else the same.
% tokenizing comments into words based on spaces between them.
% ; that is, we keep stopwords, non-lemmatized words, punctuation, etc.
% The only text transformation we perform is to tokenize comments into ``uncleaned'' words (i.e., tokens may contain punctuation etc.) by spaces such that we can obtain the bag-of-words.
% We also do not implement Term Frequency-Inverse Document Frequency (TF-IDF).
%This is because most bots operate on prompts, and only have a few variable words which they switch between comments while keeping everything else the same.
% Thus, lack of pre-processing should make distinguishing these bots easier as words are always accompanied by the same punctuation etc.
% and tend to post comments which are differentiated only to a very minor extent (if any)
% Thus, retaining representations of text exactly as it appears in comments should provide a clearer distinction between bots, who repeat similar words accompanied by the same punctuation etc., and real users.
Thus, our approach is only suitable for overt bots.
% (i.e., those attempting to mislead other users.)
% We nullify diagonal values in the similarity matrices to avoid artificially inflating the average similarity for users with fewer comments, 
We plot the Cumulative Distribution Function (CDF) of user average comment similarity in Figure~\ref{fig:cdf}.
% Once we obtain the user similarity matrix, we then nullify diagonal values (i.e., those which compare each comment to itself), and we compute the average similarity across the entire matrix.
% We nullify diagonal values to avoid artificially inflating the average similarity in smaller matrices for users with fewer comments, as these values are always the maximum of 1.
We focus on the small subset of users with similarity values $\geq 0.4$ as nearly all users are under this threshold, leaving 410 users.
% Nearly all users have a similarity under 0.4.
% Thus, we focus on the small subset of users with similarity values $\geq 0.4$, leaving 410 users.

% Following filtering based on number of comments ($> 10$) and comment similarity threshold ($\geq 0.4$), 
To determine the similarity cutoff for bots, we rely on a heuristic name-based rule as with prior work on Reddit~\cite{hurtado_bot_2019,soliman_characterization_2019}.
% which takes into account that most bots in our sample are overt, and so have bot-like names.
% For this, we rely on a simple heuristic rule to approximate an annotation procedure via which we label users as bots or not, considering that most of the bots in our sample are overt and therefore tend to have bot-like naming conventions.
We label users as bots if their usernames contain any bot-like word (bot, auto, moderator), or any popular platform name (Reddit, YouTube, Facebook, Twitter, Imgur), as many bots provide services related to these platforms and thus mention them in their names.
% (e.g., bots which comment with YouTube video descriptions under video links).
% Thus, each user is labeled as a bot or a non-bot using this heuristic approximation.
We perform a logistic regression with class weighting on user text similarity, acknowledging that some bots may not have the heuristic words in their usernames and vice-versa.
%, while some non-bots may have such words.
% We then use the model to predict the class of each of the 410 users.
We take the similarity point at which predictions switch from non-bot to bot as our cutoff (0.59).
Model predictions against heuristic annotations result in an F1 score of 0.85.
% The confusion matrix of heuristic annotations vs. model predictions is shown in Figure~\ref{fig:confusion_matrix}, which results in an F1 score of 0.85.
% \begin{figure}[t!]
%     \centering
%     \includegraphics[width=0.55\columnwidth]{figures/bot_cm.png}
%     \caption{Confusion matrix of logistic model predictions against heuristic annotations.}
%     \label{fig:confusion_matrix}
% \end{figure}

We treat any users who exceed the cutoff as bots, obtaining 111 accounts.
Further, we manually check whether the 28 users with bot-like naming conventions below the cutoff appear in bot detection subreddits (e.g., \emph{r/BotDefense}, \emph{r/BotTerminator}, etc.), state they are bots in comments/profile descriptions, or show suspicious behavior. 
% (e.g., highly similar comments).
Of these accounts, we find 26 more bots, bringing the total to 137.
% We treat the users in the two rightmost quadrants as bots as they exceed the 0.59 similarity threshold, obtaining 111 bot accounts. 
% However, we also manually check the 28 users in the bottom left quadrant as they have bot-like naming conventions.
% Specifically, we check whether the account admits to being a bot either in the comments it posts or in its profile description, whether it has been listed as a bot in bot detection subreddits (e.g., \emph{r/BotDefense}, \emph{r/BotTerminator}, etc.), or whether it shows suspicious behavior (e.g., its comments are very similar to each other with only subtle keyword differences). 
% Of these accounts, we determine that 26 are indeed bots, and we add them to our bot list for a total of 137 bots.
% We then remove any user pair in which at least one of the two accounts is one of these 137 bots.
% While we present our analyses below with the bot pairs removed, we also conduct robustness analyses with these pairs included and observe the same patterns.

\section{Readability Formulas Used}\label{app:formulas}

\begin{enumerate}
 \item \textbf{Flesch-Kincaid Grade Level:} Relies on total syllables, words, and sentences in text.
 \item \textbf{Flesch Reading Ease:} Relies on total syllables, words, and sentences in text. Uses different weighting than Flesch-Kincaid Grade Level.
 \item \textbf{SMOG Index:} Relies on number of polysyllables (i.e., words with $\geq$ 3 syllables) and number of sentences.
 \item \textbf{Coleman-Liau Index:} Relies on average number of words and average number of sentences per 100 words.
 \item \textbf{Automated Readability Index:} Relies on number of characters, words, and sentences in text.
 \item \textbf{Dale-Chall Formula:} Relies on the ratio of ``difficult'' words to total words, and the ratio of total words to total sentences. Difficult words are those that appear in a curated list.
 \item \textbf{Linsear Write Metric:} Relies on a point system based on number of syllables in each word and total number of sentences.
 \item \textbf{Gunning Fog Index:} Relies on number of words, polysyllables, and sentences.
\end{enumerate}

\section{LIWC Examples}\label{app:liwc}

We show representative comments, i.e., those that are among the highest-scoring for each LIWC category per subreddit in Table~\ref{tab:liwc}.

\begin{table*}[t]
  \centering
  \footnotesize
  \begin{tabular}{lp{5cm}p{5cm}p{5cm}}
    \toprule
      \textbf{sub} & \textbf{Authentic} & \textbf{Analytic} & \textbf{Clout} \\
      \midrule
      TD & ``Thanks man, appreciate it. I know there are some definite nut jobs out there that embarrass us quiet ones.'' & ``Worker ownership of all industries without the state is in all practical means impossible, especially in the 21st century post-industrial economy.'' & ``Facebook wants you to shut the zuck up.'' \\
      \midrule
      SFP & ``I disagree, but that's cool.'' & ``Hillary has an incredible amount of power within the establishment.'' & ``You should be ashamed of yourself.'' \\
      \midrule
      CTH & ``Where do you get your left political analysis from? Honest question.'' & ``Louisiana surreally passed a hate crime law designed to protect police.'' & ``Notice how when you’re challenged you just say more nastier things and lie?'' \\
      \midrule
      CON & ``There is some idea that reading out of a teleprompter, and saying the same old talking points, is presidential. I never understood why people felt that way.'' & ``The DNC hack certainly shows improper collaboration between the DNC and the press.'' & ``If you deny that there are liberal Republicans, you are badly out of touch.'' \\
      \bottomrule
  \end{tabular}
  \caption{Representative texts per LIWC category and subreddit.}
  \label{tab:liwc}
\end{table*}

\section{Named Entity Recognition (NER)}\label{app:ner}

For NER, we use the pre-trained \code{en\_core\_web\_trf} transformer model based on the RoBERTa architecture~\cite{liu_roberta_2019} from Python's \code{spacy} library.
To improve suitability, we annotate 5K random comments from political subreddits~\cite{rajadesingan_political_2021} and train a new model on top of the pre-trained one with an 80-10-10 train-validation-test split.
% To make the model more suitable for our use-case, 

This new model captures more informal terms (e.g., ``dems'', ``neocons'', etc.), provides increased performance against the test set (F1 score of 77.18 vs 48.62 of the pre-trained model in the annotated test set), and includes two more entity types that we added (Ideology and Website).%\footnote{We note through manual inspection that our model may be biased against longer texts as it occasionally misses entities if a multitude of them exist in a single comment; this is a limitation for a small subset of comments.}
% The patterns we report below are not substantially different from what we find using the pre-trained model.
% such that it occasionally misses entities if multiple of them exist in a comment, which is an important limitation for a subset of our texts.

\begin{figure}[t!]
    \centering
    \includegraphics[width=0.99\columnwidth]{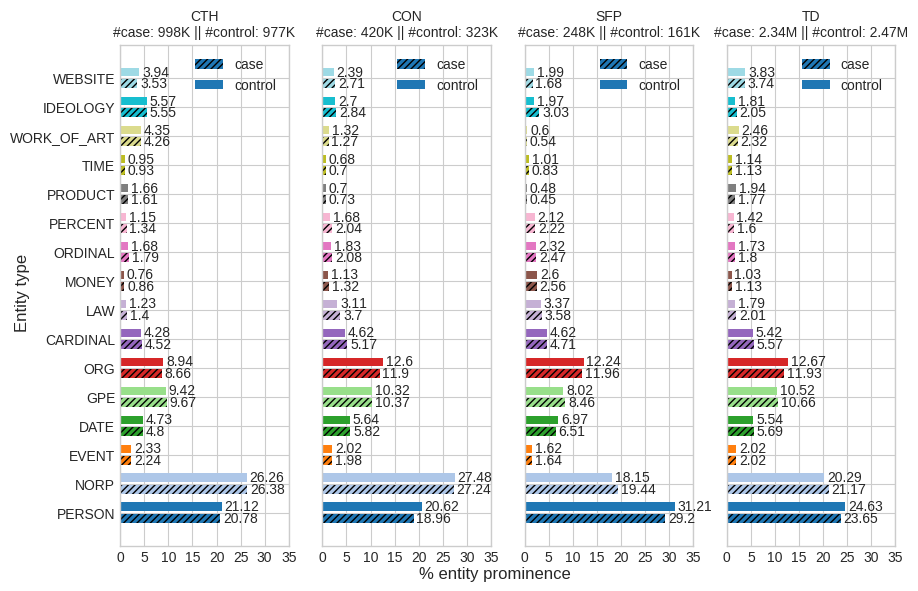}
    \caption{Prominence comparison of all entity types by subreddit group. \#case and \#control indicate total number of entities detected across all comments. Fac, Language, Quantity and Loc omitted due to $<$ 1\% prominence in all groups.}
    \label{fig:entities}
\end{figure}

We show all entity types and brief definitions in Table~\ref{tab:entity_legend}, along with any amendments we make relative to the pre-trained model.
% Note that 18 out the 20 categories are part of the pre-trained \verb|en_core_web_trf| model, however, we show the definitions we followed in our own annotation procedure which we may have slightly amended for some entities.
We perform NER on all 7.51M comments made by case and control users,\footnote{We observe similar patterns using the pre-trained model.} and plot the \emph{relative} prominence of each entity in Figure~\ref{fig:entities}.
% We observe similar patterns using the pre-trained model.

\begin{table*}[t]
  \centering
  \footnotesize
  \begin{tabular}{p{1.6cm}p{6cm}lp{6cm}}
    \toprule
       \textbf{Label} & \textbf{Definition} & \textbf{Label} & \textbf{Definition} \\
       \midrule
         \textbf{CARDINAL} & Numbers that do not fall under another type & \textbf{DATE} & Absolute or relative dates or periods \\
         \textbf{EVENT} & Named hurricanes, battles, wars, sports events, etc. & \textbf{FAC} & Facilities like buildings, airports, highways, bridges, etc. \\
         \textbf{GPE} & Geopolitical entities (countries, cities, states) & \textbf{LANGUAGE} & Any named language \\
         \textbf{ORG} & Organizations like companies, agencies, institutions, etc. & \textbf{LOC} & Non-GPE locations, mountain ranges, bodies of water \\
         \textbf{LAW$\dagger$} & Named documents made into laws, \textit{or any other official government documents like reports, proposed policy, etc.}& \textbf{NORP$\dagger$} & Nationalities or religious or political groups, \textit{or ethnic, racial, or ideological groups} \\
         \textbf{ORDINAL} & ``First'', ``second'', etc. & \textbf{MONEY} & Monetary values, including unit \\
         \textbf{PERCENT} & Percentage, including \% & \textbf{PERSON} & People, including fictional \\
         \textbf{PRODUCT} & Objects, vehicles, food, etc. (not services) & \textbf{QUANTITY} & Measurements, as of weight or distance \\
         \textbf{TIME} & Times smaller than a day & \textbf{WORK\_OF\_ART} & Titles of books, songs, movies, shows, etc. \\
         \textbf{IDEOLOGY*} & Political, economic, religious, or philosophical ideology, school of thought, or system & \textbf{WEBSITE*} & Any named website which is not referred to in the context of an organization (e.g., Reddit) \\
         \bottomrule
  \end{tabular}
  \caption{All types of entities in the model. $\dagger$The entity's definition has been amended relative to the base model, with the amendment shown in italics. *The entity was added to the trained model and does not appear in the pre-trained one.}
  \label{tab:entity_legend}
\end{table*}

% Differences between case and control users (per subreddit) are fairly subtle.
% Looking at simply which group (case or control) has the higher relative percentage of each entity, some entities are universally higher in one or the other group across all four subreddits.
Organizations and persons are mentioned more by control users than case users in all subreddits.
% (also works of art and facilities, although the differences are too small to be reliable).
Ordinal and cardinal numbers, percentages, laws, and geopolitical entities are universally mentioned more by case users.
Generally, controls refer {\em slightly} more to personified entities, while cases refer more to numeric and legal entities.
% , such as numeric entities, ratified or proposed legislation, and countries, cities, etc.
Nonetheless, these patterns are very subtle; overall, the two groups discuss similar kinds of entities.
This also holds for the exact entities mentioned per type, shown in a Google Sheet\footnote{https://docs.google.com/spreadsheets/d/e/2PACX-1vSzW-eubTn2GZpAYDFuzjHJh3mcfICoJdH7qjrBvTKqPWNsHrhr\\V44rudVGJ3RB4A/pubhtml}. % (see Table~\ref{tab:exact_entities} for more details).

\section{Follow-up Downvotes Analyses}

Here, we show complementary analyses to Section~\ref{sec:cmv_trails}.

\subsection{Early Sanctions in CMV}\label{app:early}

We consider the role of each user's \textit{first} comment in CMV as this may have a disproportionate impact~\cite{kang_motivational_2022}, especially considering the challenges faced by newcomers when entering new communities~\cite{steinmacher_systematic_2015} and the negative impact on content quality when excluding them~\cite{halfaker_dont_2011}.
We compare users who only have a single comment in the community during our observation period to those who have more than one, since the former may have left due to early perceived disapproval of their views.
% Therefore, we expect more downvotes among these first-and-last-time commenters relative to others.

We count downvoted and non-downvoted comments among these users, and compare them to 1) the total pool of comments in CMV throughout 2018 and 2) the pool of first comments by users who went on to post more.
Chi-squared tests reveal the opposite pattern: downvoted comments among one-time commenters are fewer compared to both the total pool ($\chi^2(1)$ = 131.37, \emph{p} $<$ 0.001) and the first-comment pool of other users ($\chi^2(1)$ = 295.83, \emph{p} $<$ 0.001).

This somewhat agrees with our earlier short-term analysis in that users may ``stand their ground'' in CMV.
This pattern could be due to multiple reasons; for example, users who have their first comments downvoted may be further motivated to make others see their point of view, or they may comment in highly contentious discussions, which may overall attract more engagement~\cite{davis_emotional_2021}.
% some users may just post a comment out of curiosity without any particular regard to the community's perception of their comment.

\subsection{Long-Term CMV Residents and Departees}\label{app:longterm}

\begin{figure}[t!]
    \centering
    \begin{subfigure}{0.80\columnwidth}
      \centering
      \includegraphics[width=0.80\columnwidth]{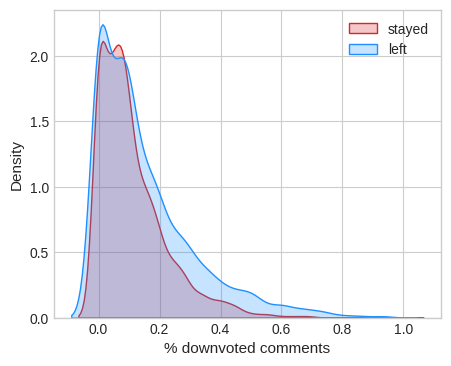}
      \caption{Base comparison}
    \end{subfigure}

    \begin{subfigure}{0.80\columnwidth}
      \centering
      \includegraphics[width=0.80\columnwidth]{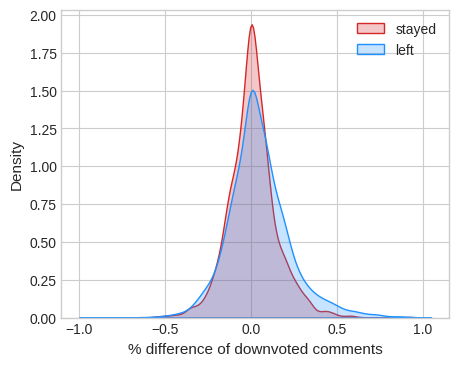}
      \caption{Comparison to home}
    \end{subfigure}
 \caption{Comparison of percentage downvoted between long-term residents and departees.}
 \label{fig:comparison}
\end{figure}

To examine whether downvotes which are consolidated and built up over the longer term play a role in whether a user leaves a discussion community, we separate users into those who stayed active in CMV throughout the year, and those who left at some point.
We sample users who posted at least 10 comments in CMV between January 1st - June 30th, 2018 (i.e., the sampling period).
From this, we subset those who also posted at least one comment in CMV between October 1st - December 31st, 2018 (i.e., the long-term residents).
The remaining subset reflects the departees.
% (i.e., difference between the initial set and the long-term residents)
The ``dead period'' (July 1st - September 30th) is to allow for consolidation of comment votes during the sampling period.

Then, we compare the proportion of \textit{downvoted} comments during the sampling period between these two sets using independent t-tests.
% If downvoting drives users out of the community, then we should expect to see a higher proportion of downvoted comments in those who left.
Indeed, we find that downvotes are more prominent among departees (\emph{N} = 4678, \emph{M} = 15\%) than long-term residents (\emph{N} = 3549, \emph{M} = 11\%), \emph{t} = 12.54, \emph{p} $<$ .001, Cohen's \emph{d} = 0.279 (small effect size), indicating that downvoting may be a factor in their eventual departure.

This pattern also holds when examining relative downvotes compared to what users receive at their homes.
We compare the \emph{difference} in percentages of downvoted comments between each user's CMV and home comments, omitting anyone who has CMV as their sole home.
% (i.e., we subtract the proportion of downvoted home comments from the proportion of downvoted r/changemyview comments over the sampling period).
% Through this, we want to observe whether users that leave the discussion community receive more relative rewards at their homes.
This confirms that departees are more downvoted in this community relative to their homes (\emph{N} = 3033, $M_{difference}$ = 4.63\%) than long-term residents (\emph{N} = 2931, $M_{difference}$ = 0.84\%), \emph{t} = 8.95, \emph{p} $<$ .001, Cohen's \emph{d} = 0.232 (small effect size).

However, looking at the distributions of each group of users with respect to downvote percentages in Figure~\ref{fig:comparison}, we see that these differences are mostly driven by the right ends of the curves, with substantial distribution overlap.
In fact, most users in both groups have no downvoted comments.
Thus, our results here are inconclusive as there may be several factors behind user departures from the subreddit.
% Thus, while downvotes do play a role in why users may not re-appear in the subreddit, there are possibly other important determinants of this behavior which we do not unpack here.

\end{document}